\documentclass[%
reprint,
superscriptaddress,
amsmath,amssymb,
aps,
prb,
]{revtex4-2}

\usepackage{graphicx}
\usepackage{dcolumn}
\usepackage{bm}
\usepackage{overpic}
\usepackage{lipsum}
\usepackage[version=4]{mhchem}
\usepackage{miller}
\usepackage{siunitx}
\usepackage{pict2e}
\usepackage{xcolor}

\let\oldput\put
\def\put(#1,#2)#3{%
	\oldput(#1,#2){\fontfamily{phv}\selectfont #3}
}

\definecolor{mattsgreen}{RGB}{0, 128, 0}
\definecolor{orange}{RGB}{254, 165, 0}
\begin{document}

\preprint{APS/123-QED}

\title{Enhanced skyrmion metastability under applied strain in \ce{FeGe}}

\author{M. T. Littlehales}
\author{L. A. Turnbull}
\affiliation{Department of Physics, Durham University, DH1 3LE, United Kingdom}
\author{M. N. Wilson}
\affiliation{Department of Physics, Durham University, DH1 3LE, United Kingdom}
\affiliation{Department of Physics and Physical Oceanography, Memorial University, A1B 3X7, Canada}
\author{M. T. Birch}
\affiliation{Max-Planck-Institut f\"ur Intelligente Systeme, Heisenbergestra{\ss}e, 70569, Stuttgart, Germany}
\author{H. Popescu}
\author{N. Jaouen}
\affiliation{Synchrotron SOLEIL, Saint Aubin, BP 48, 91192, Gif-sur-Yvette, France}
\author{J. A. T. Verezhak}
\affiliation{Department of Physics, University of Warwick, Coventry, CV4 7AL, United Kingdom}
\author{G. Balakrishnan}
\affiliation{Department of Physics, University of Warwick, Coventry, CV4 7AL, United Kingdom}
\author{P. D. Hatton}
\affiliation{Department of Physics, Durham University, DH1 3LE, United Kingdom}

\date{\today}

\begin{abstract}
Mechanical straining of skyrmion hosting materials has previously demonstrated increased phase stability through the expansion of the skyrmion equilibrium pocket. Additionally, metastable skyrmions can be generated via rapid field-cooling to form significant skyrmion populations at low temperatures. Using small-angle x-ray scattering and x-ray holographic imaging on a thermally strained 200 nm thick FeGe lamella, we observe temperature-induced strain effects on the structure and metastability of the skyrmion lattice. We find that in this sample orientation (H $\parallel$ \hkl[-110]) with no strain, metastable skyrmions produced by field cooling through the equilibrium skyrmion pocket vanish from the sample upon dropping below the well known helical reorientation temperature. However, when strain is applied along \hkl[110] axis, and this procedure is repeated, a substantial volume fraction of metastable skyrmions persist upon cooling below this temperature down to 100 K. Additionally, we observe a large number of skyrmions retained after a complete magnetic field polarity reversal, implying that the metastable energy barrier protecting skyrmions from decay is enhanced.
\end{abstract}

\maketitle

\section{Introduction}

Magnetic skyrmions, nanometric vortex-like whirls of magnetization, typically exist in chiral magnets \cite{Nagaosa2013, Rossler2006} under specific temperature and magnetic field conditions \cite{ Muhlbauer2009, Oike2016, Milde2013}. Skyrmions are normally stabilized by the competition of direct exchange, the Zeeman interaction, thermal fluctuations, and the Dzyaloshinskii-Moriya interaction (DMI), which requires a non-centrosymmetric crystal structure as present in conventional helimagnetic systems such as \ce{MnSi} \cite{Muhlbauer2009}, \ce{FeGe} \cite{Yu2011}, and the insulating ferrimagnet \ce{Cu2OSeO3} \cite{Seki2012}. Characterised by an integer winding number, skyrmions arrange in a periodic hexagonal crystal in two-dimensions, with a tube-like nature in the third dimension \cite{Birch2020, Rybakov2013}. Their structure, size and dynamical properties indicate promise for skyrmions as elements in complex computing devices such as skyrmion racetrack memory \cite{Fert_2013, Tomasello2015}, logic-gates \cite{Zhang2015}, boolean processors \cite{Paikaray_2022}, skyrmion transistors \cite{Moody2022} as well as neuromorphic \cite{Song2020}, stochastic \cite{Zhang_2020} and reservoir computing \cite{Pinna2020}.

 A number of methods have been demonstrated to increase the size of the skyrmion equilibrium pocket to lower temperatures and higher magnetic fields \cite{Chacon2015, Nii2015}. For example, reducing the thickness of the system along the applied field direction can expand the skyrmion equilibrium region by surpressing the conical phase \cite{Butenko2010, Huang2012}. Additionally, metastable skyrmion states with near-infinite lifetimes can be formed by rapid field cooling through the equilibrium skyrmion pocket to low temperatures \cite{Wilson2019, Wilson2020, Oike2016}. Although not the energy minimum of the system, metastable states with large energy barriers present a greater operating region in which skyrmions may be manipulated for potential future device applications. \\

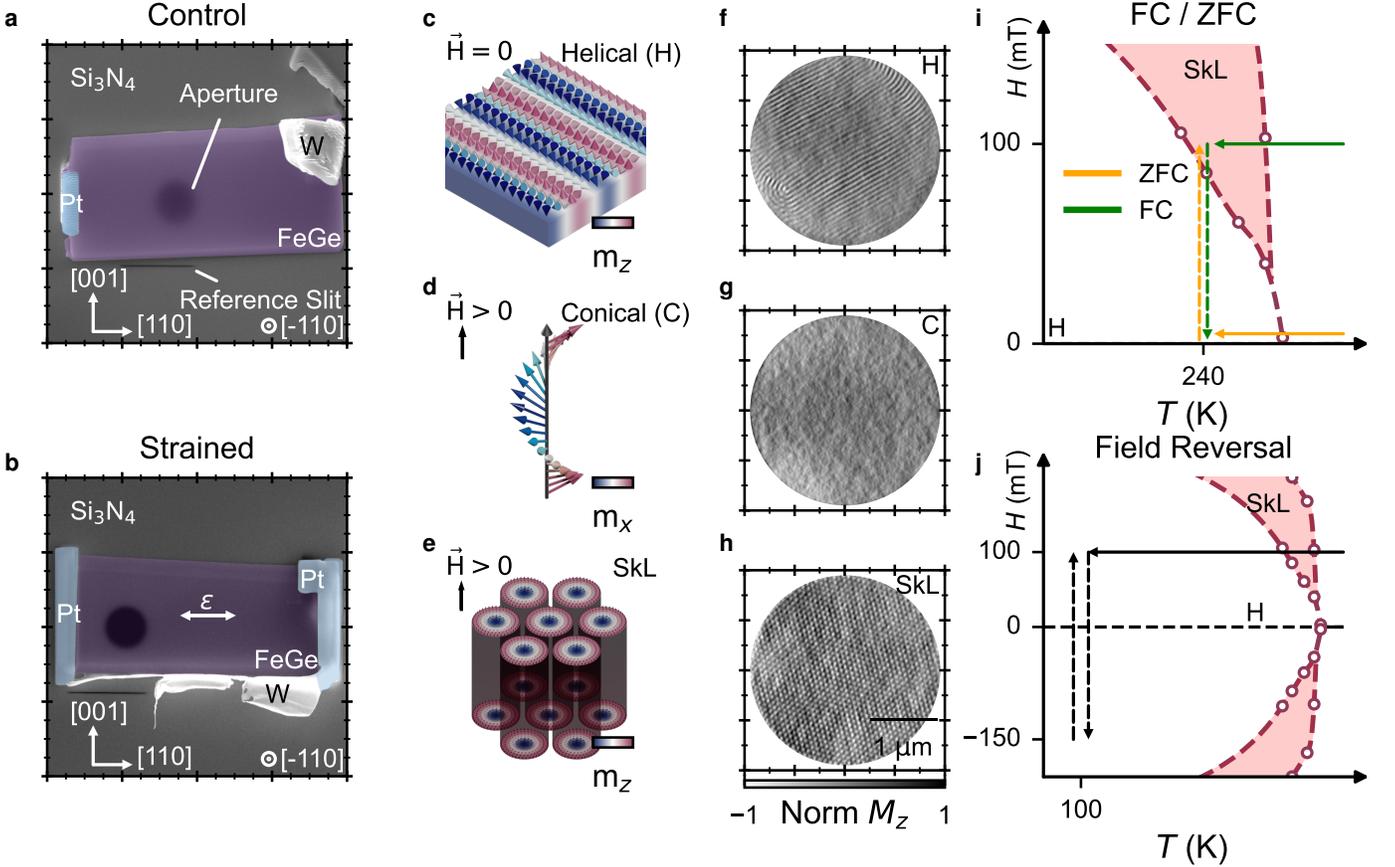
\begin{figure*}[t]
	\centering
	\begin{overpic}[trim = 0 0.5cm 0 0, width = \textwidth]{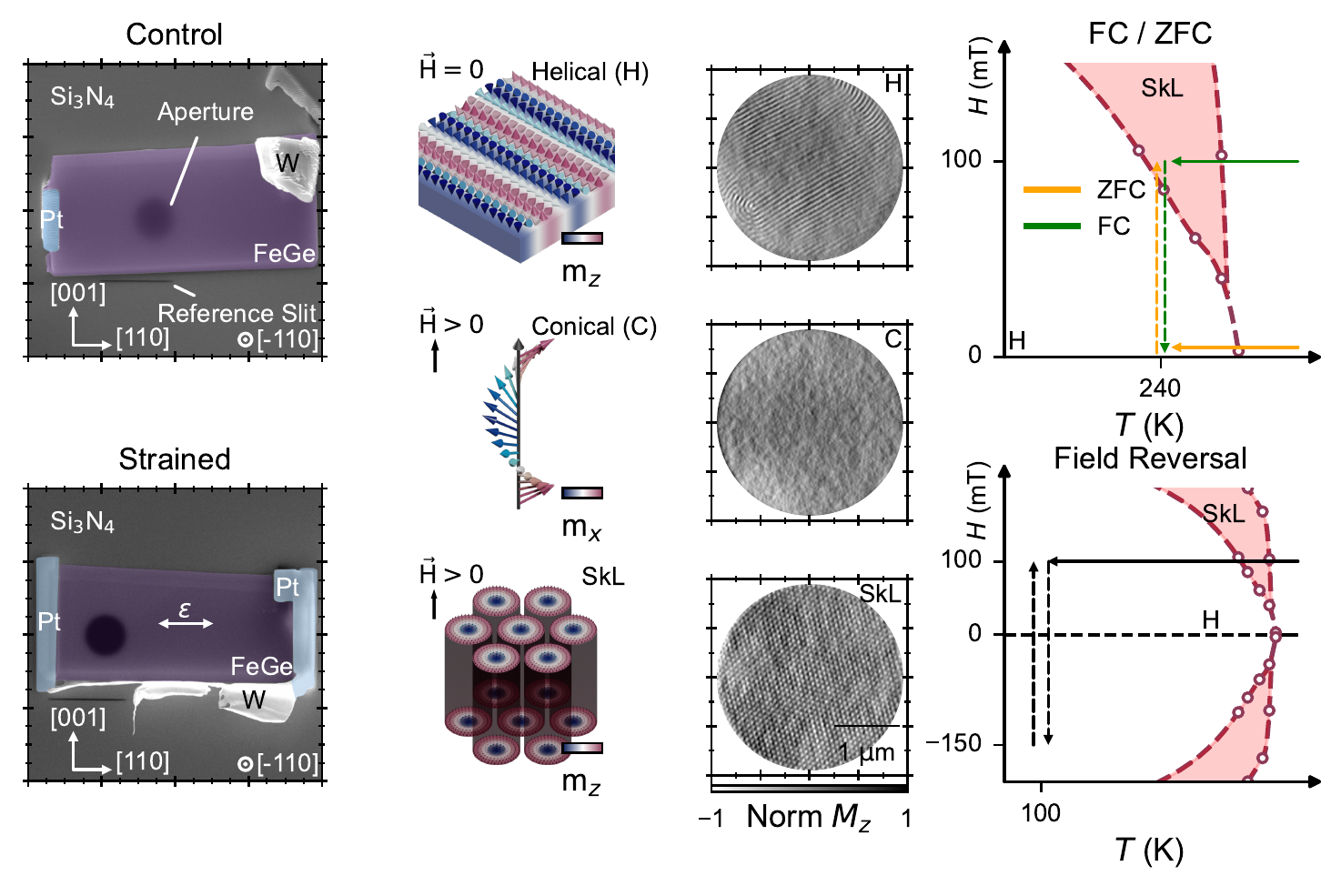}
		\put(-1,61){\textbf{a}}
		\put(-1,28){\textbf{b}}
		\put(30,61){\textbf{c}}
		\put(30,41){\textbf{d}}
		\put(30,22){\textbf{e}}
		\put(52,61){\textbf{f}}
		\put(52,41){\textbf{g}}
		\put(52,22){\textbf{h}}
		\put(71,61){\textbf{i}}
		\put(71,28){\textbf{j}}
	\end{overpic}
	\caption{\textbf{a-b} Scanning-electron-microscopy (SEM) images of control and strained lamellae respectively. Darker patches in the substrate represent a \SI{3}{\micro\meter} diameter circular aperture and \SI{6}{\micro\meter} long and \SI{15}{\nano\meter} wide reference slit milled through the substrate using focused ion beam milling. Purple indicates the \ce{FeGe} lamella, blue the platinum strips, and white the tungsten remaining from the omniprobe micromanipulator detachment. \textbf{c-e} Schematic illustrations of spin textures present within lamella, \textbf{c} Helical (H), \textbf{d} Conical (C) and \textbf{e} Skyrmion Lattice (SkL). \textbf{f-h} x-ray holographic reconstructions of spin textures in \ce{FeGe} lamellae corresponding to the illustrations in \textbf{c-e}. \textbf{i-j} Phase diagram schematics of cooling protocols used within this study. \textbf{i} zero-field-cooled (ZFC) protocol including cooling in zero-field to 240 K then increasing magnetic field to 100 mT (orange) and field-cooled protocol at 100 mT to 240 K with a reduction in magnetic field to 0 mT (green). \textbf{j} Field-reversal protocol including a field-cool at 100 mT to 100 K followed by a reversal of the magnetic field to -150 mT and back to 100 mT. }
	\label{f1}
\end{figure*}

As a near-room-temperature skyrmion host, \ce{FeGe}, is a promising candidate for the above mentioned devices. However, a major consideration in engineering novel spintronic applications lies in exploiting extrinsic effects such as mechanical strain \cite{Lei2013}. As the magnetic spin configuration in any material is fundamentally dependent on the underlying crystal structure, understanding the connection between magnetism and mechanical strain, known as magnetoelastic coupling, is paramount. Recent studies investigating strain effects on skyrmion spin textures have demonstrated anisotropic modulations of the skyrmion lattice, effectively modulating the DMI and exchange strength \cite{Shibata2015, Kang2017}. Similar anisotropic DMI has been observed in anti-skyrmion systems caused by the D$_{2\mathrm{d}}$ crystal symmetry \cite{Jena2020, Li2022}. As a consequence of these energetic changes, mechanical strain can increase the skyrmion stability region \cite{Sukhanov2019, Seki2017}, (as is also seen for applied uniaxial pressure \cite{Nii2015}) with Monte-Carlo simulations of anisotropic DMI supporting these results \cite{Hog2021, Hog2022}. Additionally, there are reports of room-temperature skyrmions existing in strain-engineered \ce{FeGe} thin films \cite{Budhathoki2020}. These results also point to a method of direct skyrmion creation, via voltage controlled application of strain with piezoelectric substrates \cite{Hu2018, Yan2019}.

In this study, we employ resonant x-ray holography \cite{Eisebitt2004, Guizar-Sicairos2007} to image the real space magnetization \cite{vanderlaan2014} of metastable skyrmion states under the effects of thermally induced tensile strain on lamella of single crystal FeGe \cite{Turnbull2021, Turnbull2022}. Through differential thermal contraction between the sample and substrate, we confirm elliptical skyrmion lattice deformation using small-angle x-ray scattering (SAXS) and demonstrate enhanced metastability of the skyrmion against helical reorientations \cite{Lebech1989, Ukleev2021, Moody2021} and magnetic field reversals. 

\section{Experimental Details}
Single crystals of \ce{FeGe} were grown via the chemical vapour transport method using 2 g of prepared FeGe powder and 2 mg/cm$^3$ of iodine transport agent with the source maintained at 450 $^\circ$C and a 50 $^\circ$C temperature gradient across the length of the furnace. After a period of 1-2 weeks, several single crystals with dimensions 1.5 x 1.5 x 1.5 mm$^3$ were obtained at the furnace cold end. Two 200 nm thick lamellae were milled from one of the single crystals with \hkl[001] and \hkl[110] directions in the plane of the sample and \hkl[-110] out of the plane of the lamella, using focused gallium ion-beam (FIB) milling. The substrate consisted of a 5$\times$5 mm, \SI{300}{\micro\meter} silicon chip with 200 nm thick \ce{Si3N4} x-ray transparent windows masked with 600 nm sputter-coated Au to avoid detector saturation and minimise x-ray background, which was then mounted on a copper holder. The lamellae were fixed to the \ce{Si3N4} windows using platinum deposition at room temperature. A \SI{3}{\micro\meter} diameter aperture and a centrally offset \SI{6}{\micro\meter} long and \SI{15}{\nano\meter} reference slit \cite{Guizar-Sicairos2007} was milled through the Au mask using FIB milling as shown in Fig. \ref{f1}a \& b.

A control sample was made by fixing a single edge of a lamella to allow free expansion and contraction with temperature (Fig. \ref{f1}a). The sample under strain was produced by attaching opposite ends as shown in Fig. \ref{f1}b. The silicon chip contracts more than \ce{FeGe} when cooling, therefore tensile strain up to a maximum of 0.25 \% is produced at 100 K along the \hkl[110] direction (See Appendix 1 for details of strain calculations).

X-ray holography was undertaken using the COMET end-station on the SEXTANTS beam line at the SOLEIL synchrotron using resonant soft x-ray scattering with both left and right circularly polarised x-rays at the Fe $L_3$ edge with an energy of 697.7 eV. This energy provided maximal magnetic contrast to absorption ratio for samples of this thickness, with the x-ray magnetic circular dichroism (XMCD) contrast isolated using a subtraction of the two circularly polarised x-ray scattering patterns. Interference patterns between the circular aperture and extended reference slit were detected on a $2048\times2048$, \SI{13.5}{\micro\meter}$^2$ pixel charge coupled device (CCD). With a sample-to-detector distance of 138 mm, this gives an effective $q$ resolution of 1 $\times10^{-3}$ nm$^{-1}$ at this energy corresponding to a real-space pixel size of 24.4 nm.

The samples were cooled to a base temperature of 100 K via a liquid helium cryostat. Magnetic fields up to a maximum of 150 mT were applied using a four-pillar permanent magnet array. Schematics of helical, conical and skyrmion phases are shown in Fig. \ref{f1}c-e with corresponding x-ray holograms shown in Fig. \ref{f1}f-h.  At low magnetic fields, we observe helical states, consisting of continuous rotations of the magnetic spins around a propagation vector as shown in Fig. \ref{f1}d. Under finite magnetic fields applied along the incident x-ray beam direction, the helical propagation vector rotates to align along the magnetic field direction, while each spin gains a component along the field direction (Fig. \ref{f1}d). Since the component of magnetization along the beam direction is constant, no variation in the magnetic contrast is observed in Fig. \ref{f1}g. Under specific temperatures and magnetic fields, we observe magnetic skyrmions as a six-fold arrangement of light contrast circles surrounded by a dark background as shown in Fig. \ref{f1}h.

To discern possible differences in the strain effects on metastable and equilibrium skyrmions, we employed three distinct field-cooling and sweeping procedures. In the zero-field cooling (ZFC) and field-cooling (FC) regimes, the samples were cooled from above the Curie temperature ($T\mathrm{_C}$ $\approx$ 278 K) \cite{Yu2011} under applied fields of 0 and 100 mT respectively to a target temperature of 240 K. The ZFC/FC procedures were followed by increasing/decreasing magnetic field sweeps at 240 K, with holographic images collected at regular intervals. A third procedure involved FC to 100 K followed by a magnetic field-reversal, denoted here as FR, where the field was decreased to zero, the orientation switched with respect to the beam direction, then increased. Each procedure is denoted in the phase diagrams in Fig. \ref{f1}i \& j with solid arrows indicating cooling procedures and dashed arrows showing magnetic field changes. Metastable skyrmions were only formed during FC protocols, as the field-cooling history passed through the equilibrium skyrmion region.  

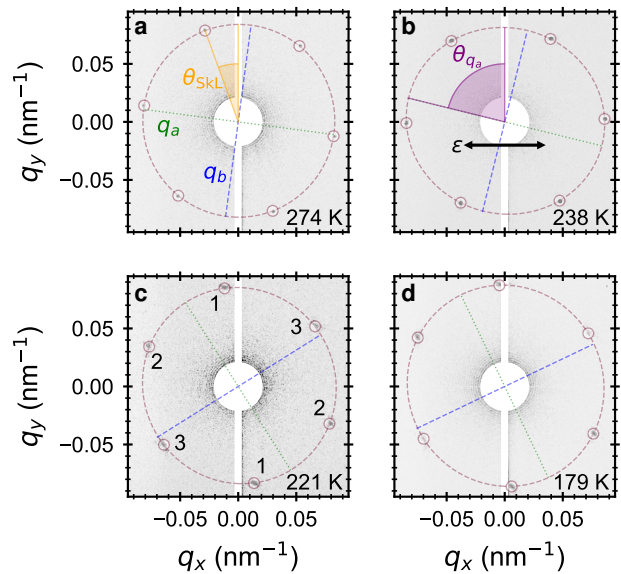
\begin{figure}[t]
	\begin{overpic}[width=0.45\textwidth]{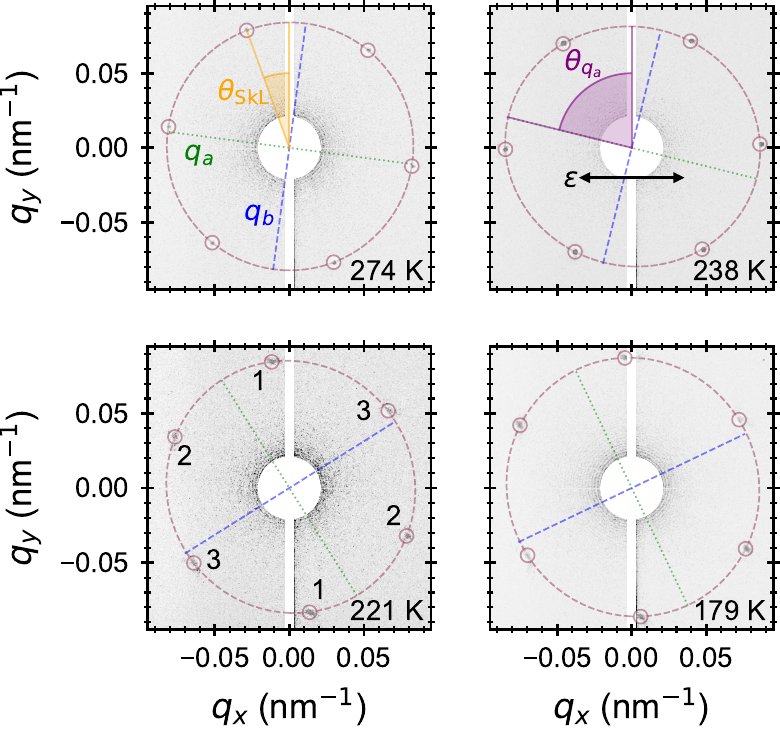}
		\put(20,89){\textbf{a}}
		\put(64,89){\textbf{b}}
		\put(20,45){\textbf{c}}
		\put(64,45){\textbf{d}}
	\end{overpic}
	\caption{\textbf{a-d} SAXS patterns of the strained FeGe lamellae at various temperatures. Red dashed line shown in \textbf{a} represents an elliptical fit to the magnetic scattering peaks, green dotted and blue dashed lines correspond to the semi-major ($q_a$) and semi-minor axes ($q_b$) respectively. Skyrmion lattice orientation, $\theta_{\mathrm{SkL}}$, and ellipse orientation, $\theta_{q_a}$, are measured anticlockwise from the vertical shown by orange lines in \textbf{a} and purple lines in \textbf{b}. The strain direction in reciprocal space is represented by $\varepsilon$ in \textbf{b}. Pairs of skyrmion peaks are labelled in \textbf{c} corresponding to the analysis in Fig. \ref{f2c}a.}
	\label{f2}
\end{figure}

\section{Results and discussion}
To accurately measure mechanical strain effects on magnetic skyrmions and associated magnetic textures, we utilised complementary real and reciprocal space techniques, small-angle x-ray scattering (SAXS) and x-ray holography, allowing us to quantify both skyrmion distortion and enhanced metastability.

We first investigate strain-induced skyrmion lattice distortion using SAXS. By field cooling at 100 mT through the skyrmion pocket we formed metastable skyrmions at low temperatures while inducing increased lamella strain with decreasing temperature, due to differential thermal contraction.  Magnetic scattering, corresponding to the interaction between circularly polarized x-rays and the $m_z$ component of the sample magnetization, is shown in reciprocal space as a map of the $q_x$, $q_y$ plane. Consequently, characteristic six-fold skyrmion scattering patterns \cite{Muhlbauer2009, Wilson2020} are observed down to a base temperature of 100 K with a corresponding strain of 0.25 \% as seen in Fig. \ref{f2}a-d. Within the equilibrium pocket, the estimated strain at 274 K (Fig. \ref{f2}a) is 0.03 \% corresponding to an oblateness of f = 0.01: effectively an isotropic skyrmion lattice (Fig. \ref{f1}h). 

As a function of increasing strain, we observed elliptical distortion of the skyrmion lattice, alongside alignment of the skyrmion reciprocal space peaks along the strain direction (Fig. \ref{f2}b), indicating strain dependence on both the structure and orientation of the skyrmion lattice. By fitting an ellipse to the six magnetic satellites, we can characterise both the direction of the distortion in reciprocal space and it's magnitude in the form of oblateness,

\begin{equation}
    f = \frac{q_a - q_b}{q_a}, 
\end{equation} 
where $q_a$ and $q_b$ correspond to the reciprocal length of the semi-major and semi-minor axes of the ellipse respectively (See Fig. \ref{f2}a for example). 

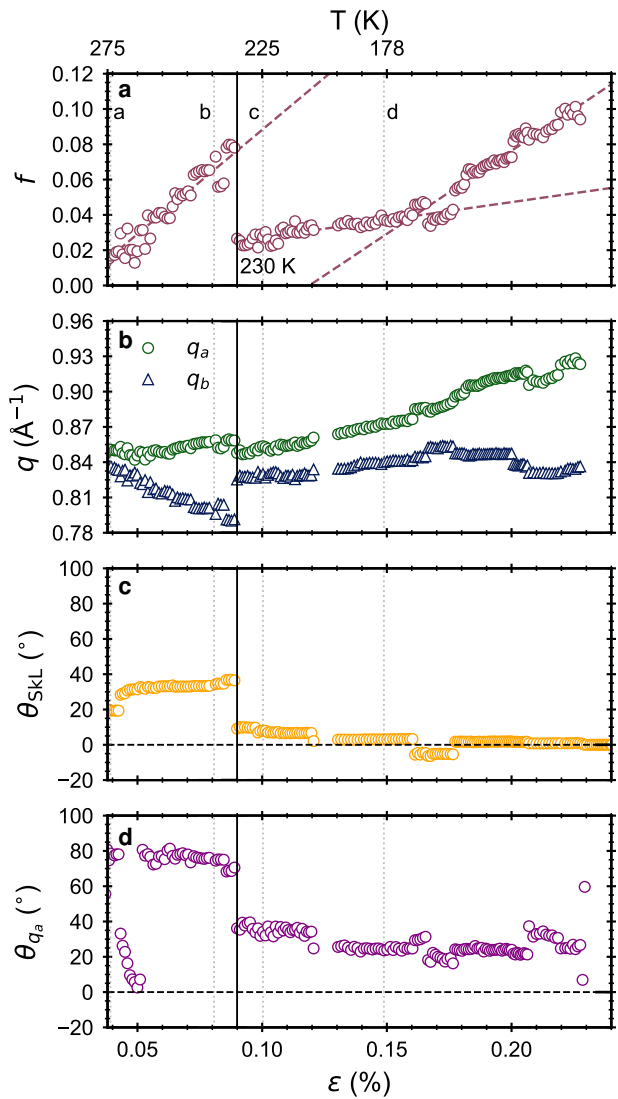
\begin{figure}
	\begin{overpic}[width=0.45\textwidth]{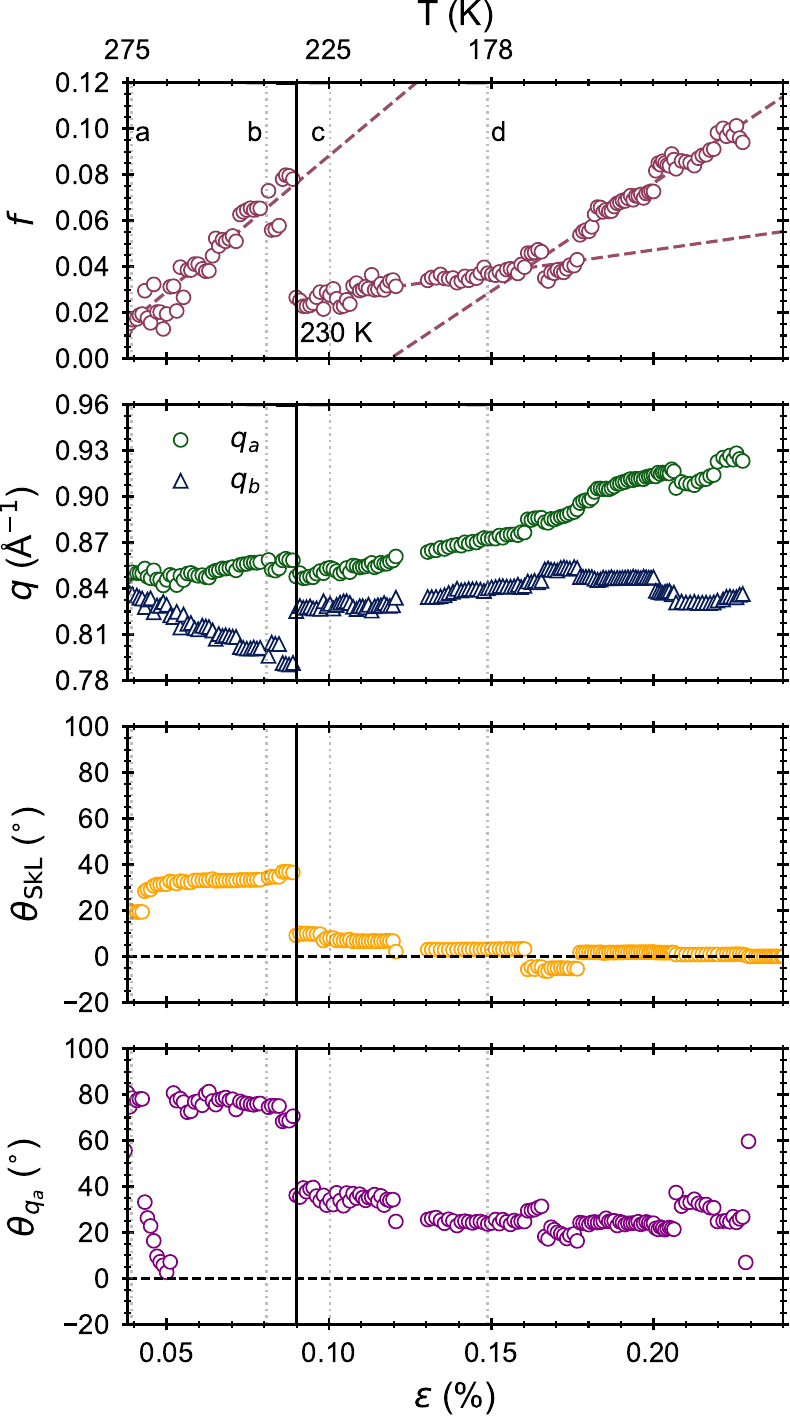}
		\put(10,92){\textbf{a}}
		\put(10,69){\textbf{b}}
		\put(10,46.5){\textbf{c}}
		\put(10,23.5){\textbf{d}}
	\end{overpic}
	\caption{Parameters measured during 100 mT field-cooled strained lamella. \textbf{a} Fitted ellipse oblateness as a function of increasing strain. Dashed lines indicate straight line fits to linear regions. \textbf{b} Semi-major (green circle) and semi-minor (blue triangle) $q$-vectors, $q_a$ and $q_b$ respectively, of the fitted ellipses as a function of increasing strain. \textbf{c} Skyrmion lattice orientation, $\theta_{\mathrm{SkL}}$ versus strain. \textbf{d} Fitted ellipse orientation, $\theta_{q_a}$ versus strain. Intrinsic reorientation of the magnetocrystalline anisotropy easy axes indicated by black vertical line at 230 K on each plot. Temperatures of SAXS patterns in Fig. \ref{f2}\textbf{a-d} are represented by labelled grey vertical dotted lines in panels \textbf{a-d} and labelled in panel \textbf{a}.}
	\label{f2b}
\end{figure}

Figure \ref{f2b}a shows the resulting skyrmion oblateness as a function of increasing strain. We observe a linearly increasing distortion from room temperature to 230 K as shown in Fig. \ref{f2b}a, at which point the distortion relaxes. Additionally, the reduction of $q_b$ between 275 K and 230 K (Fig. \ref{f2b}b) is indicative of the intrinsic helical winding length increasing along the \hkl[100] direction (See Fig. \ref{f2}b), perpendicular to the strain. The divergence of $q_a$ and $q_b$ under increasing tensile strain has previously been referred to as the skyrmion lattice ``Poisson ratio" \cite{Kang2017,Hu2019} and indicates that the magnetic state inherits the anisotropy of the crystal lattice distortion. Shibata \textit{et al.,} have previously proposed the possibility of strain-induced anisotropic modulation to the DMI to account for this behaviour \cite{Shibata2015}, as the intrinsic helical winding length is described by $\lambda = 2\pi D/J$ where $D$ is the DMI parameter and $J$ is the exchange interaction strength. They argue that strain-induced modification to the exchange interaction is not sufficient to account for the magnitude of the anisotropy. Additionally, supporting first principles calculations indicate that $D$ is enhanced, and hence $q$ is suppressed \cite{Koretsune2015}, matching our observations.

At 230 K, the reduction in skyrmion lattice oblateness coincides with a distinct reorientation of the lattice (Fig. \ref{f2b}c), best represented by determining the angle from vertical, anticlockwise to the nearest magnetic peak as represented by the orange angle in Fig. \ref{f2}a. In \ce{FeGe}, at approximately this temperature, the magnetocrystalline anisotropy easy axes reorient from the \hkl<100> directions to the \hkl<111> \cite{Lebech1989, Ukleev2021}, in this case, causing a reduction of the lattice distortion and a rotation of the skyrmion lattice. Increasing the strain shows a linear increase in lattice oblateness with a different gradient to the distortion from 275-230 K, as shown by the linear fits in Fig. \ref{f2b}a, inferring that the energy term provided by strain is dependent on the crystallographic directions and subsequently, the skyrmion orientation. This is supported by a further, smaller reorientation at $\approx$ 160 K, where the distortion gradient changes again and we see a similar divergence of $q_{a,b}$ to the higher temperature effects (Fig. \ref{f2b}b), further evidence for a positive skyrmion Poisson ratio in \ce{FeGe}. In addition, the orientation of the fitted ellipse as plotted in Fig. \ref{f2b}d shows a reorientation of the distortion direction from roughly the strain direction (allowing for some strain along the \hkl[001] direction), to approximately 54.7$^\circ$ from the strain axis. This is identical to the reorientation of the easy axes from \hkl[100] to \hkl[111], inferring that there is a dependence of the DMI anisotropy direction on the magnetocrystalline easy axes.

\begin{figure}
	\begin{overpic}[width=0.45\textwidth]{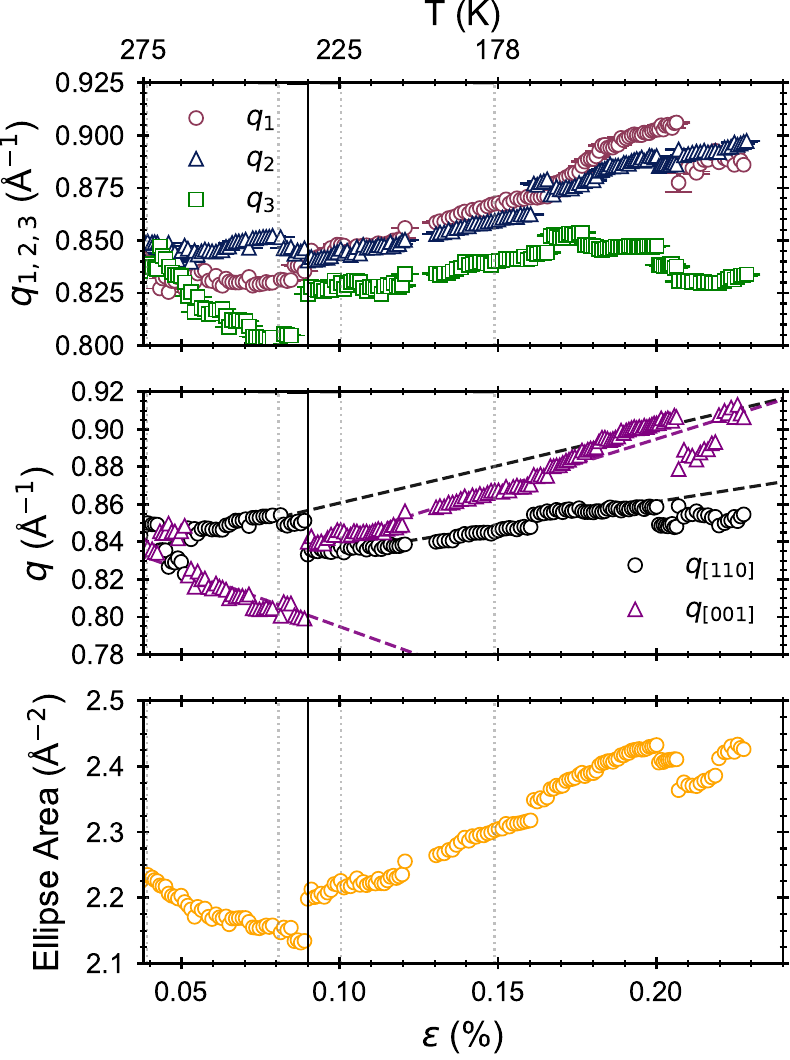}
		\put(15,89){\textbf{a}}
		\put(15,60){\textbf{b}}
		\put(15,31){\textbf{c}}
	\end{overpic}
	\caption{\textbf{a} $q_{1,2,3}$ vectors defined in Fig. \ref{f2}c plotted as a function of increasing strain. \textbf{b} $q$ vectors parallel to \hkl[110] (black circles) and \hkl[001] (purple triangles) versus strain. \textbf{c} Ellipse area calculated using Area = $\pi q_aq_b$ versus strain. Intrinsic reorientation of the magnetocrystalline anisotropy easy axes indicated by black vertical line at 230 K on each plot. Temperatures of SAXS patterns in Fig. \ref{f2}\textbf{a-d} are represented by labelled grey vertical dotted lines in panels \textbf{a-d} and labelled in panel \textbf{a}.}
	\label{f2c}
\end{figure}

In Fig. \ref{f2c}a, we plot the $q$-vectors for each of the pairs of peaks as defined in Fig. \ref{f2}c. We observe an immediate oblateness in the ellipse at 275 K as shown in Fig. \ref{f2}a due to a reduced $q_1$ vector, with a component orthogonal to the strain direction. With increasing strain, both $q_1$ and $q_2$ remain approximately constant, with $q_3$ linearly reducing until the easy axis reorientation at 230 K. After this reorientation, we observe all $q$-vectors increase with reducing temperature, consistent with previous studies \cite{Wilson2020}, with slightly different gradients. The strain-dependence of $q$ along the crystallographic axes shown in Fig. \ref{f2c}a shows that along the strain direction ($\parallel$ \hkl[110]), there is a linear increase in $q$, the gradient of which is similar both before and after the critical reorientation. In contrast, perpendicular to the strain direction ($\parallel$ \hkl[001]), we observe a steep decrease in $q$ before the reorientation, corresponding to an enhancement in DMI in agreement to theoretical studies \cite{Koretsune2015}. After the critical reorientation, $q\parallel$\hkl[001] linearly increases with a steeper gradient than along the strain direction, inferring that DMI is diminished after the reorientation, providing further evidence for magnetocrystalline anisotropy directionality dependence on the strain-induced DMI anisotropy.

These results are further reflected in Fig. \ref{f2c}c, where we observe a slight reduction in the total area of the ellipse with increasing strain followed by a large linear increase in the area after the easy axis reorientation. This is indicative that there is a change in skyrmion size as a consequence of DMI anisotropy when there is a full volume fraction constrained through topological protection of skyrmion number. After the reorientation, the system loses this constraint as a number of skyrmions are destroyed, allowing a reorientation of the skyrmions along a preferred direction, and arrange at their intrinsic spacing, dependent on (anisotropic) $D$ and $J$. 

\begin{figure}[b]
	\begin{overpic}[width=0.4\textwidth]{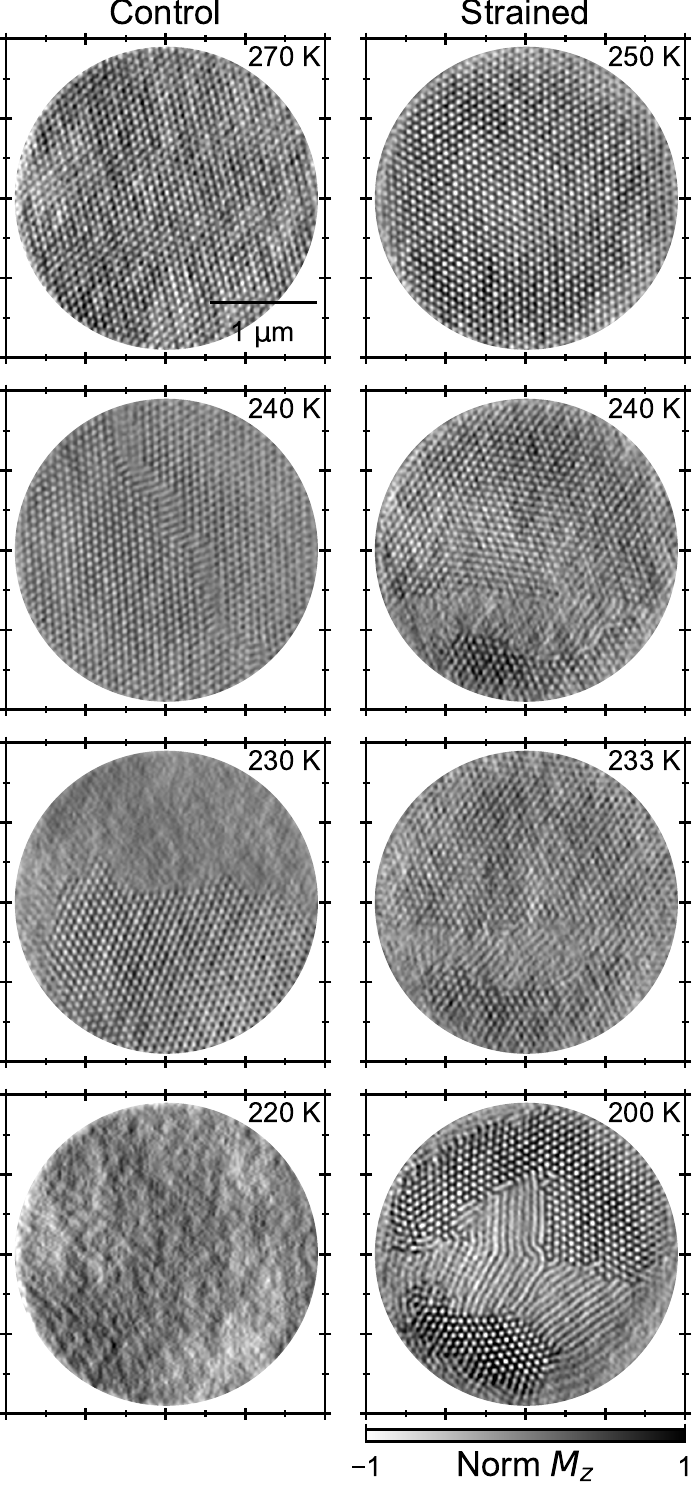}
		\put(1,95.5){\textbf{a}}
		\put(1,71.5){\textbf{b}}
		\put(1,48){\textbf{c}}
		\put(1,24){\textbf{d}}
		\put(25.5,95.5){\textbf{e}}
		\put(25.5,71.5){\textbf{f}}
		\put(25.5,48){\textbf{g}}
		\put(25.5,24){\textbf{h}}
	\end{overpic}
	\caption{\textbf{a-d} X-Ray holographic images of skyrmion states in an unstrained lamella formed through a field cooling process at 100 mT from 270 K to 220 K. \textbf{e-h} Corresponding holographic images of a strained lamella through field cooling at 100 mT from 270 K to 200 K.}
	\label{f3}
\end{figure}

\begin{figure*}
	\begin{overpic}[width = 0.8\textwidth]{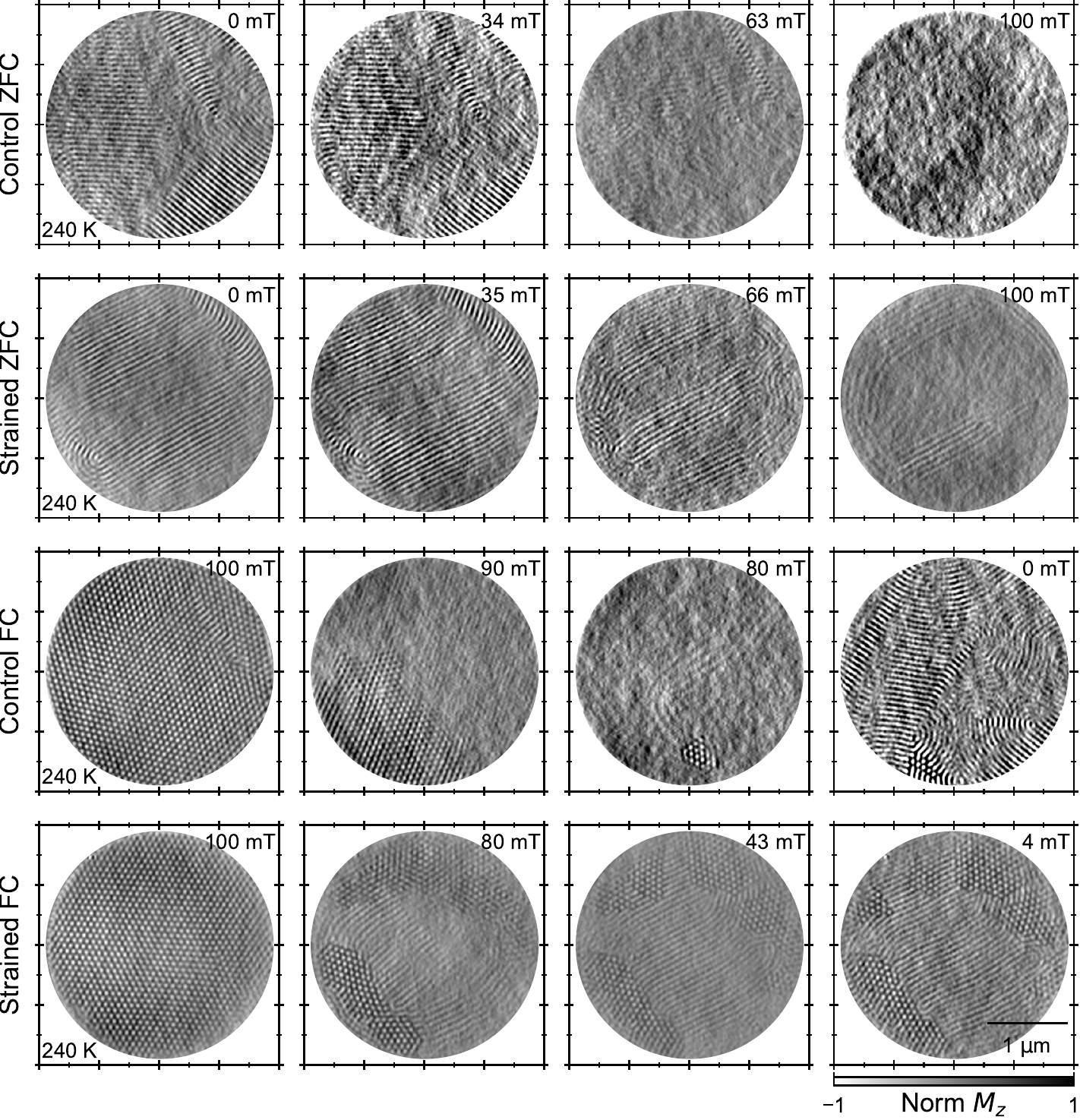}
		\put(0,100){\textbf{a}}
		\put(4,97.5){\textbf{i}}
		\put(27.5,97.5){\textbf{ii}}
		\put(51,97.5){\textbf{iii}}
		\put(75,97.5){\textbf{iv}}
		\put(0,76){\textbf{b}}
		\put(4,73){\textbf{i}}
		\put(27.5,73){\textbf{ii}}
		\put(51,73){\textbf{iii}}
		\put(75,73){\textbf{iv}}
		\put(0,51.25){\textbf{c}}
		\put(4,48.5){\textbf{i}}
		\put(27.5,48.5){\textbf{ii}}
		\put(51,48.5){\textbf{iii}}
		\put(75,48.5){\textbf{iv}}
		\put(0,27){\textbf{d}}
		\put(4,24){\textbf{i}}
		\put(27.5,24){\textbf{ii}}
		\put(51,24){\textbf{iii}}
		\put(75,24){\textbf{iv}}
		\linethickness{1pt}
		\put(5,101){\color{orange}\vector(85,0){85}}\textbf{}
		\put(5,76.75){\color{orange}\vector(85,0){85}}\textbf{}
		\put(5,52){\color{mattsgreen}\vector(85,0){85}}\textbf{}
		\put(5,27.75){\color{mattsgreen}\vector(85,0){85}}\textbf{}
	\end{overpic}
	\caption{\textbf{a} Holographic images of the ZFC protocol in the control sample measured at 240 K, \textbf{b} corresponding images taken in the strained sample after the identical procedure. \textbf{c-d} Images of FC sweeps in control and strained sample respectively. Orange arrows indicate the increase in magnetic field after ZFC, green solid arrows indicate reduction in magnetic field after FC.}
	\label{f4}
\end{figure*}
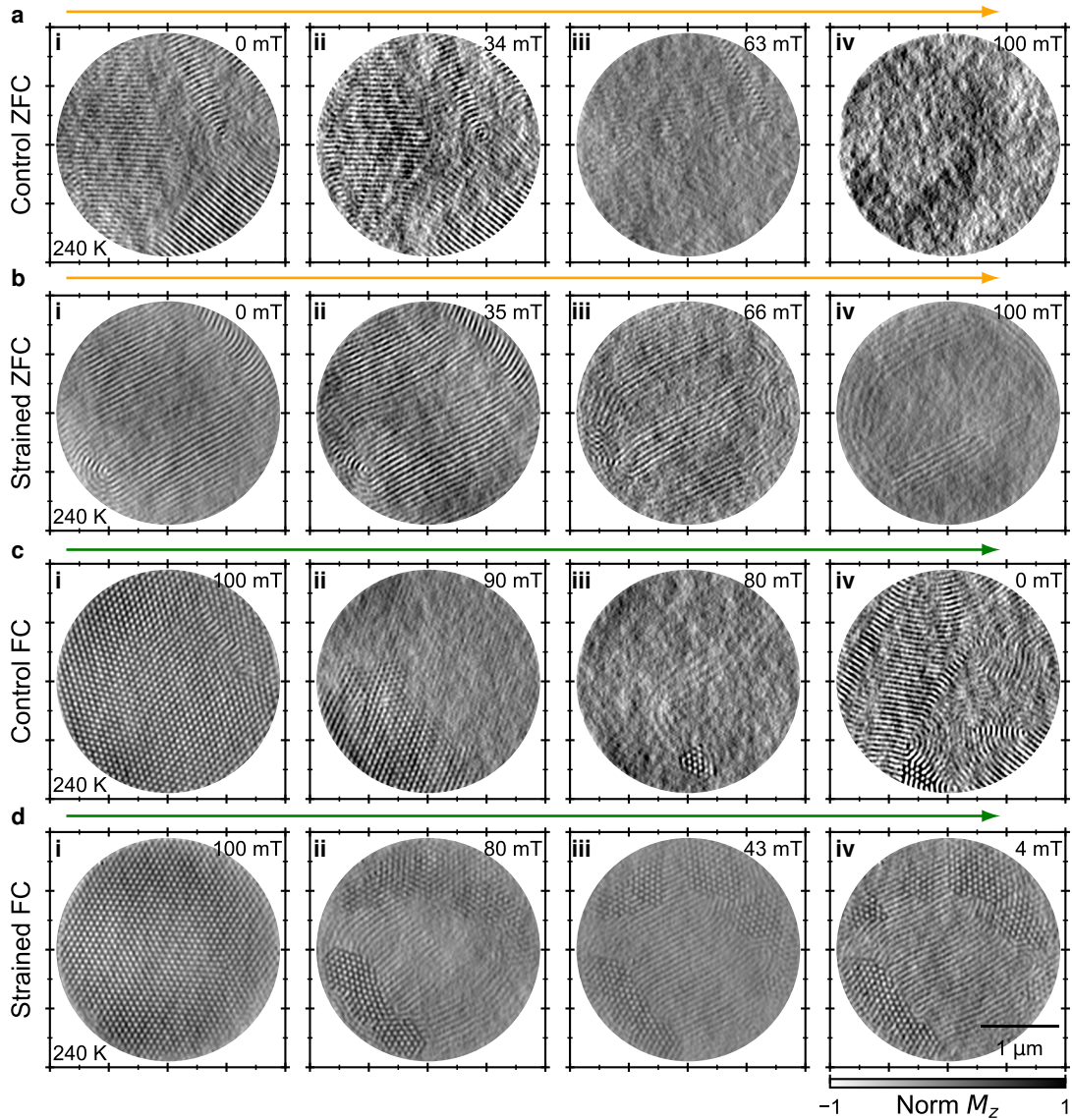

Within this experimental geometry we were unable to decouple strain and temperature effects. The coupling between temperature and strain causes a maximum oblateness of 0.1, found at 100 K, 0.25 \% strain, with a number of reductions in the maximum oblateness caused be temperature induced reorientation. Within the freely expanding control sample, we observed a minimal amount of skyrmion lattice distortion up to an oblateness of 0.04 at 230 K (See S.I. [URL will be inserted by publisher]). This is attributed to a nominal amount of constriction formed by attaching one side of the lamella to the substrate. However, the large difference in the skyrmion distortion indicates that these results can be attributed to the straining of the lamella. In addition, we observed significant differences in the evolution and stability of the skyrmion lattice with respect to the helical reorientations and captured this using x-ray holography.

We performed holographic imaging of both samples when following a similar 100 mT field cooling procedure from above $T\mathrm{_C}$ to 200 K. The images reveal the different evolution of volume fraction of skyrmions in the control and strained samples, as exhibited in Fig. \ref{f3}a-h. Interestingly, in the control lamella at 220 K, the skyrmion lattice suffers a complete volume fraction loss as shown in Fig. \ref{f3}d, indicating that the intrinsic easy-axis realignment provides enough energy to overcome the metastable energy barrier, in turn acting to unwind the skyrmion lattice. In comparison, skyrmions are retained within the strained lamella which only suffers small volume fraction reductions as shown in Fig. \ref{f3}e-g, to such an extent that at 200 K, the skyrmion lattice is still present coexisting with a complex helical domain (Fig. \ref{f3}h). The intrinsic easy axis reorientation is therefore expected to raise the energy of the skyrmion state, thus forcing a transition to the competing conical phase. However, with the addition of strain, the metastable energy barrier is increased such that the easy axis reorientation does not provide enough energy to overcome the skyrmion to cone transition, therefore, the skyrmions remain metastable.

Previous computational studies on strain in skyrmion systems suggests that the magnetoelastic coupling induced anisotropic DMI acts to stabilise A-phase skyrmions to higher magnetic fields and lower temperatures \cite{Hog2021, Hog2022}. To test this prediction, we performed ZFC and FC protocols as previously described and shown in Fig. \ref{f1}i \& j. In the control ZFC at 240 K (Fig. \ref{f4}a(i-iv)), with increasing magnetic field, we observe the ground state helical phase at 0 mT (Fig. \ref{f4}a(i)) reduce in intensity until 100 mT where zero magnetic contrast corresponds to a conical phase with the wavevector aligned along the field direction, parallel to the incident beam direction. The corresponding ZFC sweep in the strained sample (Fig. \ref{f4}b(i-iv)) indicates a similar phase evolution however at 100 mT, with a reduced volume fraction, the helical state still remains. This shows that both that the previous field-cooled states in Fig. \ref{f3} were metastable and that the helical phase is stable up to higher out-of-plane magnetic fields. 

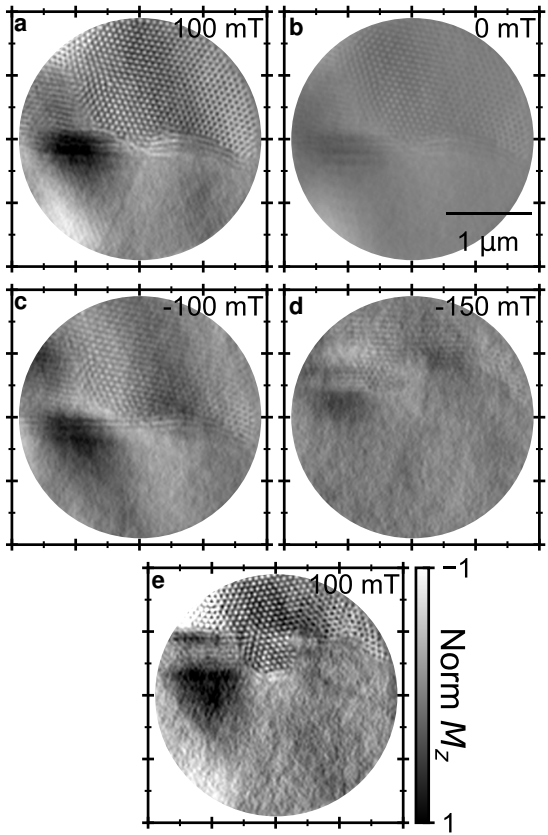
\begin{figure}[b]
	\begin{overpic}[width=0.4\textwidth]{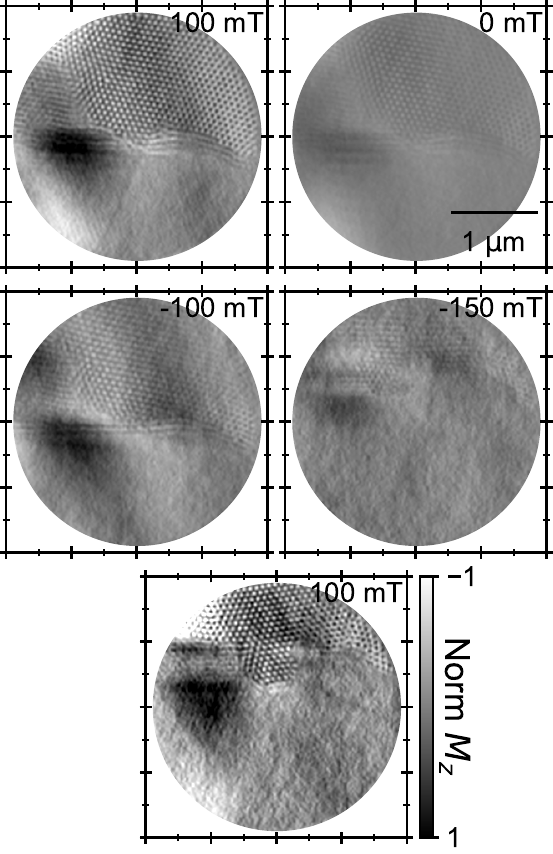}
		\put(1,97){\textbf{a}}
		\put(34,96.5){\textbf{b}}
		\put(1,63.5){\textbf{c}}
		\put(34,63){\textbf{d}}
		\put(17.5,30){\textbf{e}}
	\end{overpic}
	\caption{\textbf{a-e} X-ray holographic images taken during FR protocol indicated in Fig. \ref{f1}j. Magnetic states are shown after a 100 mT field-cool to 100 K, measured at 100 mT \textbf{a}, 0 mT \textbf{b}, reversing the direction of the magnetic field to -150 mT \textbf{c}, -150 mT \textbf{d} then back to the initial state at 100 mT \textbf{e}.}
	\label{f5}
\end{figure}

Fig. \ref{f4}c(i-iv) shows a magnetic field sweep from 100 mT to 0 mT after FC to 240 K to produce metastable skyrmions in the control sample. We observed significant volume fraction loss at 90 mT followed by only a small raft of skyrmions existing at 80 mT (Fig. \ref{f4}c(iii)). Finally at 0 mT, we observed a multidomain helical phase with six skyrmions present in the bottom of the aperture. Similar rafts of skyrmions have previously been observed by Tang. \textit{et. al.} \cite{Tang_2021} through magnetic field reversals, where up to 56 skyrmions were created within a bundle. Comparing to the strained case after FC, we observe over 200 skyrmions persisting after dropping the field to 0 mT, with the most significant volume fraction loss observed in the initial field change to 80 mT (Fig. \ref{f4}d(ii)). It can be seen that although a large number of skyrmions are lost after the initial drop in field, a greater number are retained to the lower magnetic fields within the strained sample. At this temperature, the calculated tensile strain of 0.06\% causes an increase in the metastable energy barrier, thus enhancing the skyrmion metastability. 

To investigate the extent of the enhanced metastability with increasing strain, we formed metastable skyrmions at 100 K by field-cooling at 100 mT, as indicated in Fig. \ref{f1}j, leading to a tensile strain of 0.25\%. This formed an initial state shown in Fig. \ref{f5}a with skyrmions present in the top half of the aperture, and a conical state in the bottom half, separated by a small boundary region of helical domains. Fig. \ref{f5}a-e show a magnetic FR loop, dropping to negative fields, and returning back to positive fields all while remaining at 100 K. Uncharacteristically, the skyrmions exist at all fields and are retained even after the magnetic field reversal. This is reminiscent of the skyrmion bundles observed by Tang \textit{et. al.}, \cite{Tang_2021} found after a magnetic FR in unstrained, 150 nm thick, \ce{FeGe} lamella \cite{Tang_2021}. Our results show that a much larger number of skyrmions are retained as a consequence of unaxial strain, potentially indicating that mechanical strain can be used to stabilise larger skyrmion bundles or other higher order topological objects. Similar effects were not present within the control sample as the skyrmions decayed immediately as the field was reduced (See S.I [URL will be inserted by publisher]). Similarly, ZFC to 100 K and sweeping the magnetic field up to 100 mT in the control sample showed no signs of skyrmions, indicating that the effects observed can be attributed to strain effects on metastable magnetic phases. \\

\section{Conclusions}
In conclusion, we have investigated mechanical strain effects on equilibrium and metastable skyrmions in FeGe lamellae using SAXS and x-ray holography. The evolution of a metastable skyrmion lattice as a function of decreasing temperature and increasing applied strain was tracked in order to observe effects on the structure and stability. We confirm that applied strain elliptically distorts the metastable skyrmion lattice and increases the metastability, allowing the field-cooled metastable lattice to withstand the well known magnetic easy axis switching from the \hkl<100> to the \hkl<111> directions. In contrast, non-strained metastable skyrmions are destroyed in the easy axis switching process, inferring that strain acts to increase the metastable energy barrier. Furthermore, the application of tensile strain at 100 K causes the metastable skyrmion lattice to survive against a magnetic field polarity flip, with skyrmions surviving to high magnetic fields parallel to their core. Similarly, helices under tensile strain are maintained to higher magnetic fields thus enhancing the helical to conical energy barrier, potentially providing a method of improving the stability of skyrmions, skyrmion bundles and other higher order topological states.  

\begin{acknowledgements}
We acknowledge SOLEIL Synchrotron for our time on the SEXTANTS Beamline under proposal ID 20201531. This work was supported by the UK Skyrmion Project EPSRC Programme Grant (EP/N032128/1). M. T. Littlehales acknowledges the support of the Science and Technology Facilities Council (STFC) and the ISIS Neutron and Muon Source. \\
\end{acknowledgements}

\appendix

\section{Strain Estimation}
\begin{figure}
	\begin{overpic}[trim = 3 0 0 0, width = 0.4\textwidth, percent]{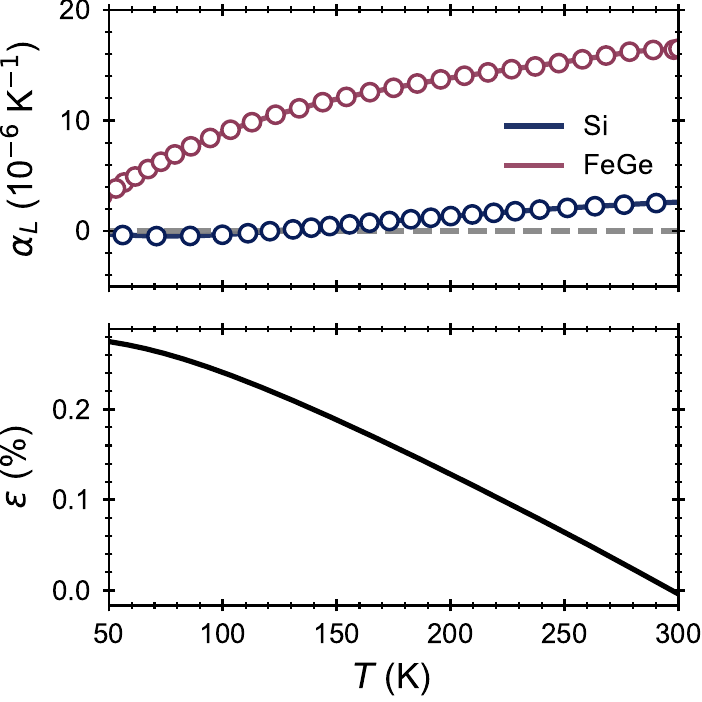}
		\put(0,100){\textbf{a}}
		\put(0,50){\textbf{b}}
	\end{overpic}
	\caption{a) Thermal expansivities of \ce{Si} (black) and FeGe (red) versus temperature \cite{Wilhelm_2016, lyon1977}. b) Estimated strain calulated using Equation \ref{Eq:strain}.}
	\label{sup_f1}
\end{figure}

The expected strain as a function of temperature is estimated using a simple model based on the differences in the thermal expansion coefficients of FeGe \cite{Wilhelm_2016} and Si \cite{lyon1977} plotted in Fig. \ref{sup_f1}a. Assuming that the strain is uniaxial in the direction perpendicular to the constrained edges as shown in Fig. \ref{f1}j, the length with respect to temperature is given by the linear thermal expansivity:
\begin{equation}
	\alpha_L = \frac{1}{L_0}\frac{dL}{dT} \Rightarrow L(T) = L_0\int_{T_i}^{T_f}\alpha_L(T)dT.
\end{equation}
Where $L_0$ corresponds to the original length of the sample, $\alpha_L(T)$ to the temperature dependent thermal expansivity and $T_{i,f}$ to the initial and final temperature respectively. Assuming fixed boundary conditions at the sample edges enables strain calculation simplifications. This assumes that the substrate is unaffected by the thermal expansion in the FeGe lamella, and that the final length of the lamella at any temperature is determined by the thermal contraction of the substrate alone. Consequently, the estimated strain is as follows:
\begin{align}
	\varepsilon = \frac{\Delta L}{L} &= \frac{L_\text{Substrate}(T) - L_\text{FeGe}(T)}{L_\text{FeGe}(T)} \\ &= \frac{\int_{T_i}^{T_f}\alpha_\text{Substrate}(T)dT}{\int_{T_i}^{T_f}\alpha_\text{FeGe}(T)dT} - 1.
	\label{Eq:strain}
\end{align}
The \ce{Si3N4} substrate is mounted on a silicon plate much larger than the substrate itself. Therefore, the thermal expansion of the substrate here is dependent upon the thermal expansion of the underlying silicon plate. Subsequently, it can be assumed that the substrate terms in Eq. \ref{Eq:strain} are relative to the silicon mounting plate rather than the \ce{Si3N4} membrane. The estimated strain as a function of temperature is plotted in red in Fig. \ref{sup_f1}b. At 100 K a strain percentage of approximately 0.25\% is expected as previously reported \cite{Shibata2015,Seki2017,Wang2018}. \\

\newpage 
\bibliography{strain_fege.bib}

\end{document}